\documentclass[sigconf]{acmart}
\usepackage{soul, color}

\usepackage{graphicx}
\usepackage{subcaption} % recommended for subfigures
\usepackage{booktabs}
\usepackage{multirow}
\usepackage{arydshln} % For dotted lines
\usepackage{amsmath} % For \pm
\usepackage{caption}
\usepackage{dblfloatfix} % fixes placement of figure*

\AtBeginDocument{%
  }

\setcopyright{acmlicensed}
\copyrightyear{2018}
\acmYear{2018}
\acmDOI{XXXXXXX.XXXXXXX}
\acmConference[Conference acronym 'XX]{Make sure to enter the correct
  conference title from your rights confirmation email}{June 03--05,
  2018}{Woodstock, NY}
\acmISBN{978-1-4503-XXXX-X/2018/06}

\begin{document}

%%
%% The "title" command has an optional parameter,
%% allowing the author to define a "short title" to be used in page headers.
\title{ DySTAN: Joint Modeling of Sedentary Activity and Social Context from Smartphone Sensors}

%{DySTAN: Dynamic Cross-Stitch and Attention for Multi-Task Recognition of Sedentary and Social Contexts in the Wild Using Smartphone Sensors}

%%
%% The "author" command and its associated commands are used to define
%% the authors and their affiliations.
%% Of note is the shared affiliation of the first two authors, and the
%% "authornote" and "authornotemark" commands
%% used to denote shared contribution to the research.
\author{Aditya Sneh}
%\email{adityas19@iiserb.ac.in}
\affiliation{%
  \institution{IISER Bhopal}
  \city{}
  \state{}
  \country{India}
}

\author{Nilesh Kumar Sahu}
%\email{nilesh21@iiserb.ac.in}
\affiliation{%
  \institution{IISER Bhopal}
  \city{}
  \state{}
  \country{India}
}

\author{Snehil Gupta}
%\email{nilesh21@iiserb.ac.in}
\affiliation{%
  \institution{AIIMS Bhopal}
  \city{}
  \state{}
  \country{India}
}

\author{Haroon R. Lone}
%\email{haroon@iiserb.ac.in}
\affiliation{%
  \institution{IISER Bhopal}
  \city{}
  \state{}
  \country{India}
}

%\renewcommand{\shortauthors}{Trovato et al.}

%%
%% The abstract is a short summary of the work to be presented in the
%% article.
\begin{abstract}
Accurately recognizing human context from smartphone sensor data remains a significant challenge, especially in \textit{sedentary settings} where activities such as studying, attending lectures, relaxing, and eating exhibit highly similar inertial patterns. Furthermore, \textit{social context}—whether a person is alone or with someone—plays a critical role in understanding user behavior, yet is often overlooked in mobile sensing research. To address these gaps, we introduce \textit{LogMe}, a mobile sensing application that passively collects smartphone sensor data (accelerometer, gyroscope, magnetometer, and rotation vector) and prompts users for hourly self-reports capturing both sedentary activity and social context. Using this dual-label dataset, we propose \textit{DySTAN} (Dynamic Cross-Stitch with Task Attention Network), a multi-task learning framework that jointly classifies both context dimensions from shared sensor inputs. It integrates task-specific layers with cross-task attention to model subtle distinctions effectively. DySTAN improves sedentary activity macro F1 scores by 21.8\% over a single-task CNN-BiLSTM-GRU (CBG) model and by 8.2\% over the strongest multi-task baseline, Sluice Network (SN). These results demonstrate the importance of modeling multiple, co-occurring context dimensions to improve the accuracy and robustness of mobile context recognition. 
%To support future research, we release the DySTAN source code publicly\footnote{\url{https://drive.google.com/drive/folders/1DihG5XsBjI_DbJpDQ1387rdWVrcR-4Sv?usp=sharing}}.

\end{abstract}



%%
%% Keywords. The author(s) should pick words that accurately describe
%% the work being presented. Separate the keywords with commas.
%% A "teaser" image appears between the author and affiliation
%% information and the body of the document, and typically spans the
%% page.

\received{20 February 2007}
\received[revised]{12 March 2009}
\received[accepted]{5 June 2009}

%%
%% This command processes the author and affiliation and title
%% information and builds the first part of the formatted document.
\maketitle

\section{Introduction}
Students frequently engage in sedentary behaviors for prolonged durations—often between eight to ten hours daily—which poses serious health risks including obesity, poor mental well-being, and increased vulnerability to anxiety and depression \cite{huang2022screen, shi2023associations, lee2018effect, babaeer2022physical, castro2018correlates}. Mobile sensing technologies offer promising avenues for passively detecting such behaviors and enabling timely, personalized interventions to promote healthier lifestyles in real time.

\begin{figure}[]
    \centering
    \begin{subfigure}{0.32\columnwidth}
        \includegraphics[width=\linewidth, keepaspectratio]{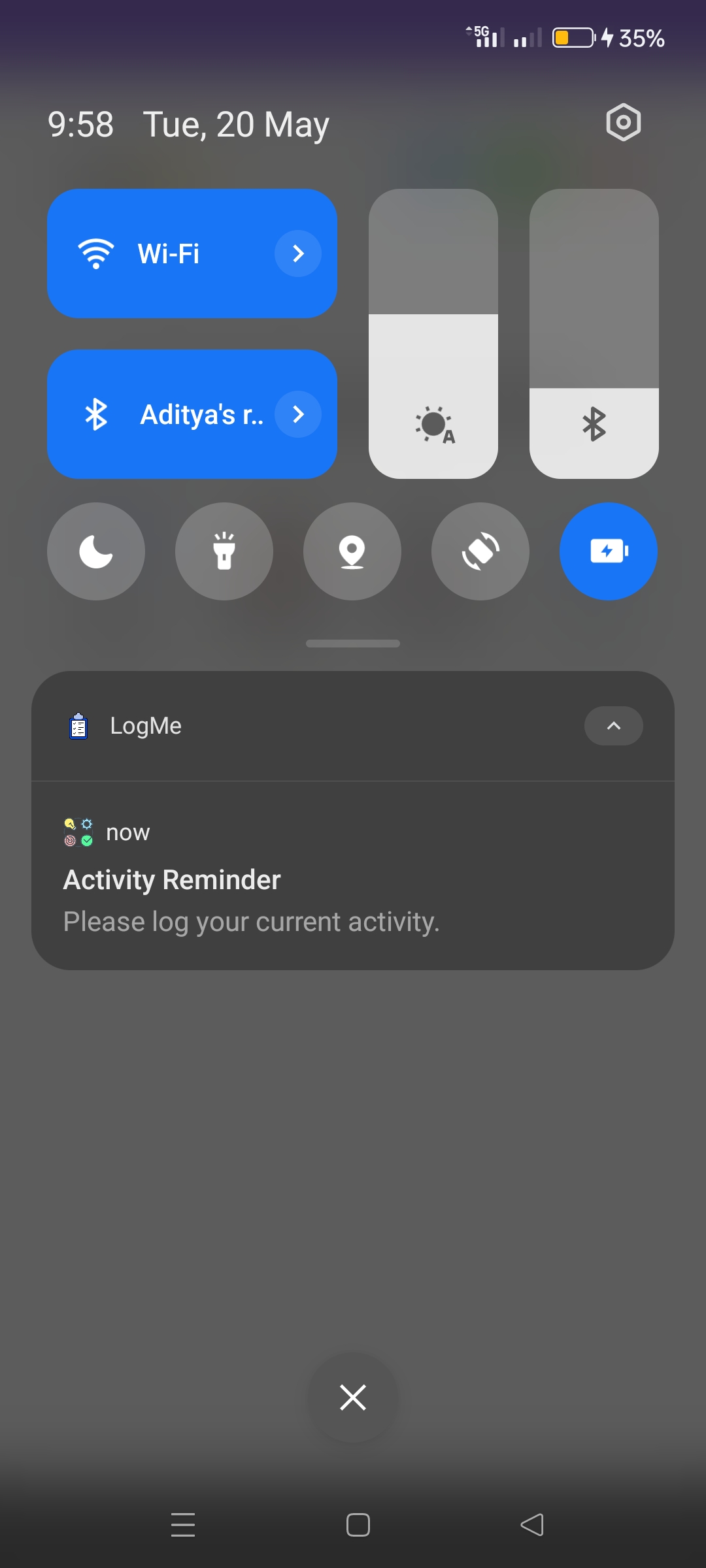}
        \caption{Hourly \textit{LogMe} reminder.}
        \label{fig:activity_context}
    \end{subfigure}
    \hfill
    \begin{subfigure}{0.32\columnwidth}
        \includegraphics[width=\linewidth, keepaspectratio]{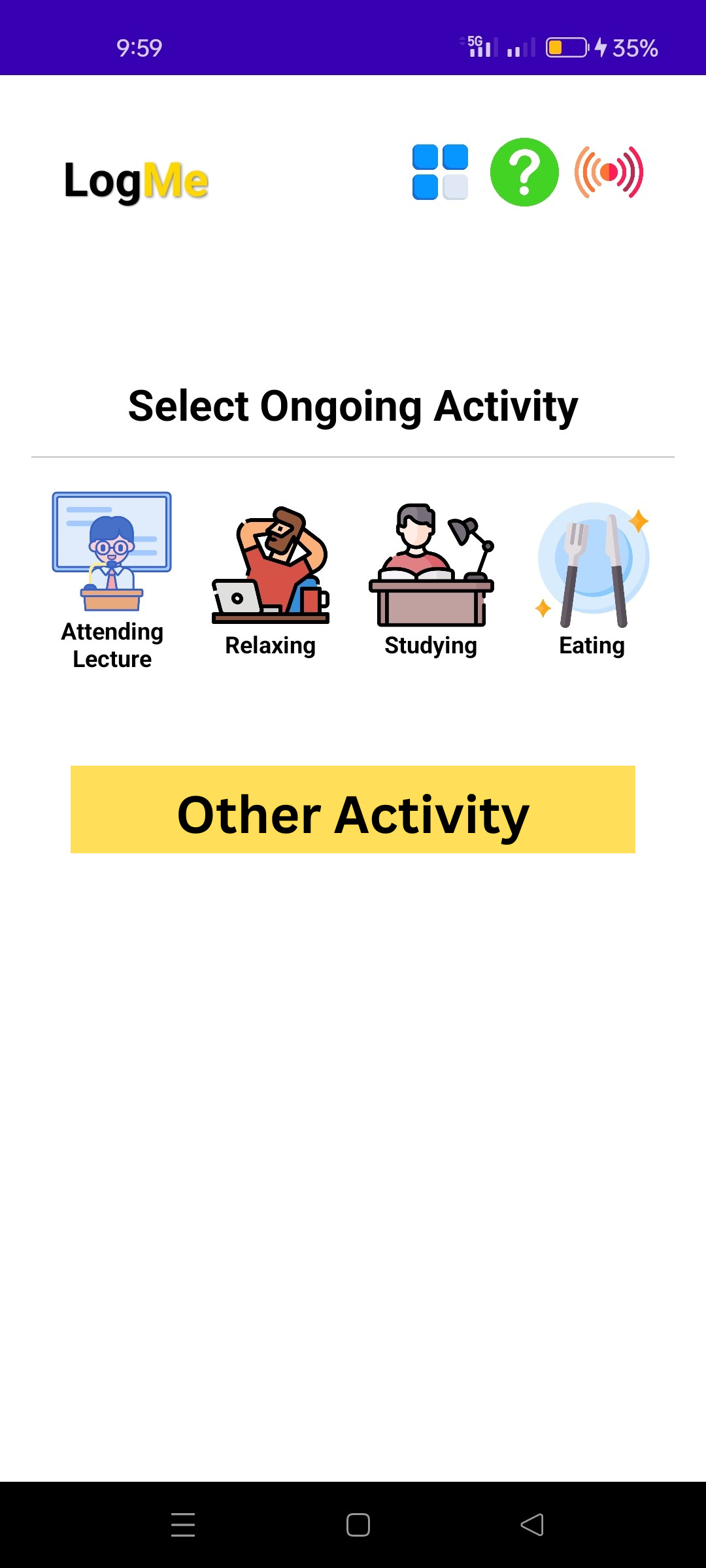}
        \caption{Sedentary activity selection.}
        \label{fig:social_context}
    \end{subfigure}
    \hfill
    \begin{subfigure}{0.32\columnwidth}
        \includegraphics[width=\linewidth, keepaspectratio]{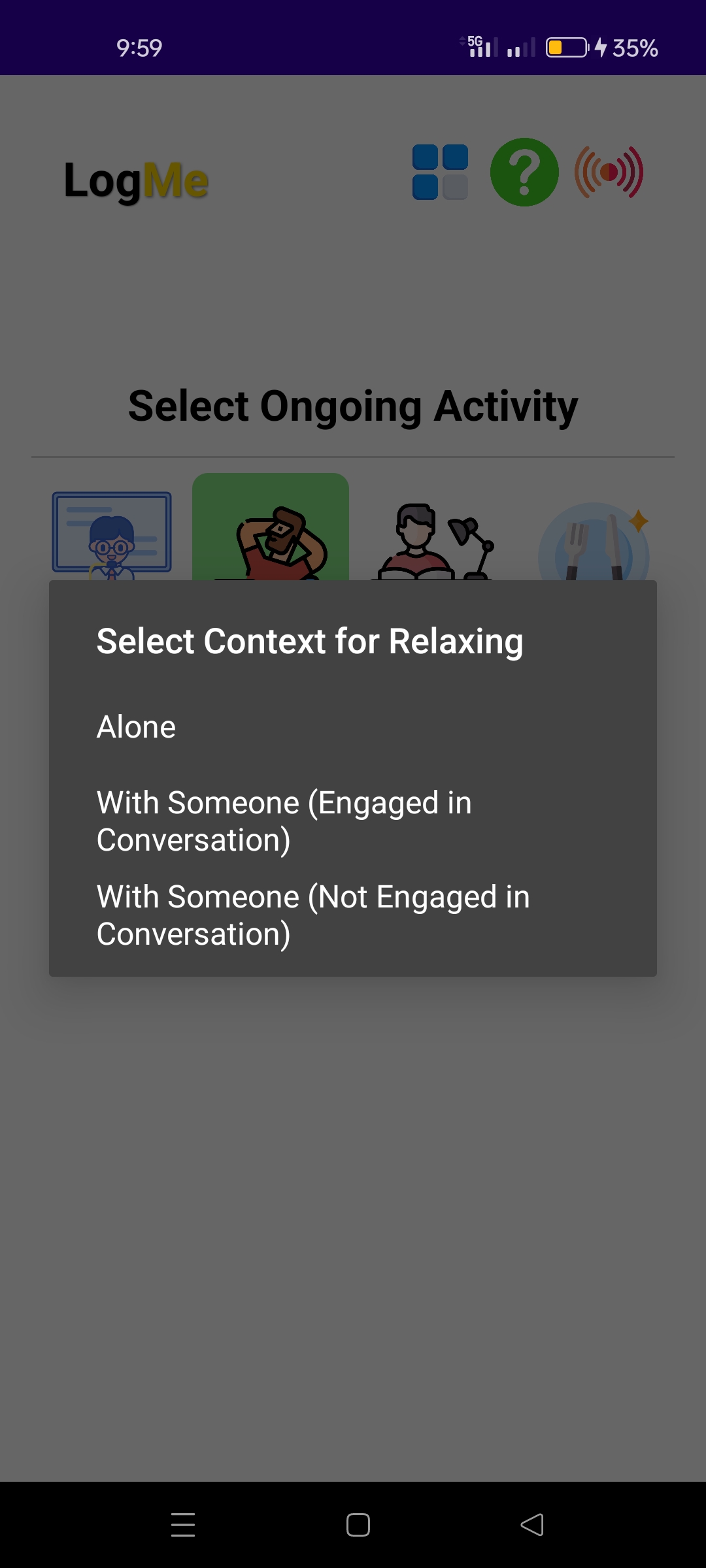}
        \caption{Social context selection.}
        \label{fig:intervention_prompt}
    \end{subfigure}

    \caption{\textit{LogMe} application flow.} 
    \Description{Screenshots showing the user interaction flow in the LogMe application, including activity context selection, social context selection, and intervention prompt.}
    \label{fig:logme_flow}
\end{figure}

The widespread use and acceptance of smartphones among students has led to increasing interest in mobile health (mHealth) solutions tailored to the student population. However, delivering contextually appropriate interventions remains challenging. Many sedentary activities—such as attending lectures, studying, relaxing, or eating—exhibit similar patterns in inertial sensor data, making it difficult to distinguish among them solely from movement signals \cite{sinha2021smartphone, fahim2018context}. Moreover, interventions that are helpful in one social context (e.g., relaxing alone) may be disruptive in others (e.g., studying or dining socially), highlighting the need for accurate detection of both \textit{activity type} and \textit{social context}~\cite{mader2024learning}.
In particular, social context—whether a person is alone or with someone—plays a key role in shaping intervention acceptance \cite{nahum2018just, mishra2021detecting}. For example, prompting someone during solitary relaxation may be appropriate, but interrupting during a group meal or conversation may reduce receptivity or even cause annoyance.

Despite its importance, social context remains underexplored in current mobile sensing models. Most existing approaches either ignore overlapping and co-occurring contexts (e.g., eating while conversing), or lack multi-label annotations, which hinders comprehensive recognition \cite{assi2023complex}. As a result, many models struggle to generalize to real-world scenarios where contexts blend fluidly. These limitations underscore the need for richer, multi-context recognition models that can jointly infer \textit{sedentary activities} and \textit{social settings} from passive smartphone sensing.

To address these gaps, we developed \textit{LogMe}, a mobile sensing app that collects passive Inertial Measurement Unit (IMU) sensor data along with active self-reports of sedentary activity—Attending Lecture (AL), Relaxing (R), Studying (S), Eating (E)—and social context—Alone (A), With Someone (Engaged in Conversation) [WSEIC], With Someone (Not Engaged in Conversation) [WSNEIC]. Using this dataset, we introduce \textit{DySTAN}, a multi-task learning model that jointly predicts both contexts from shared sensor inputs. DySTAN effectively captures task-specific distinctions and consistently outperforms existing methods. Following are our key contributions:

\begin{itemize}
%\item \textbf{Multi-task Model – DySTAN:} We propose and release \textit{DySTAN}~\cite{our-model}, a multi-task neural network that jointly infers sedentary activity and social context from shared sensor inputs. It consistently outperforms single-task and traditional multi-class baselines across multiple evaluation metrics. [HAS MODEL LINK]
\item \textbf{Multi-task Model – DySTAN:} We propose and release \textit{DySTAN}, a multi-task neural network that jointly infers sedentary activity and social context from shared sensor inputs. It consistently outperforms single-task and traditional multi-class baselines across multiple evaluation metrics.

\item \textbf{Sedentary and Social Context Dataset (SSCD):} We introduce a real-world dataset comprising 50 Hz IMU sensor streams (accelerometer, gyroscope, magnetometer, and rotation vector) collected from 57 university students over a two-week period. Each session is annotated with both sedentary activity and social context labels, making this one of the few datasets to support fine-grained, multi-label recognition in naturalistic settings.

\item \textbf{Mobile Sensing Platform – LogMe:} We develop \textit{LogMe}, an Android app that passively collects sensor data while prompting users for brief, context-rich self-reports. It enables  dual-label data collection in ecologically valid settings.
\end{itemize}

\section{Related Work}

% Early Human Activity Recognition (HAR) using inertial measurement unit (IMU) data showed strong results under controlled conditions—for instance, boosted decision trees combining phone and smartwatch data achieved a macro-F1 of about 82.9\% across six activities, with static postures recognized above 90\% accuracy \cite{thottempudi2024high,lateef2022human,kuo2023human,padmanabha2025egocharm,stoeve2021laboratory}, and an SVM using tri-axial accelerometer, gyroscope, and magnetometer data distinguished five subtle sitting poses with near-perfect accuracy (99.9\%) \cite{tang2021upper}. However, recognizing sedentary behaviors ``in the wild'' remains challenging: generic models for daily activities such as studying, eating, and attending lectures typically achieve around 0.70 AUROC, improving to 0.79--0.89 with country-specific adaptation. 

Early research on smartphone-based inertial sensing achieved high accuracy for basic activities such as running, walking, and stair climbing. However, classification performance drops significantly for fine-grained sedentary contexts such as sitting and standing, even with advanced deep learning models. For example, Haghi et al.~\cite{haghi2023recognizing} showed that while a hybrid CNN-LSTM model achieved up to 98\% accuracy for gross activities, performance for static postures like sitting and standing was notably lower, with F1 scores around 75--94\%. Similarly, Salice et al.~\cite{salice2024nonintrusive} reviewed that even state-of-the-art non-intrusive sensing systems often struggle to robustly differentiate subtle postures, with accuracy frequently dropping below 85\% for fine-grained sedentary states.

In parallel, researchers have explored social context inference. Classical classifiers distinguishing ``alone'' versus ``with others'' often exceed 90\% AUC~\cite{mader2024learning}; for example, Mäder et al.\ report $>$90\% AUC for companionship status, while Kammoun et al.\ achieve approximately 75\% AUC for eating companionship using gradient boosting~\cite{kammoun2023understanding}. However, these models typically ignore co-occurring contexts---such as eating while conversing---and struggle with poor classification performance on complex social situations. Moreover, both classical machine learning and deep learning models generally treat activity and social inference as separate tasks, highlighting the need for unified modeling approaches~\cite{arshad2022human,liu2011survey}.

To overcome these limitations, multi-task learning (MTL) has been increasingly applied to smartphone sensing. MTL approaches leverage shared feature layers to model related tasks, improving classification accuracy jointly. For example, Azadi et al.~\cite{azadi2024robust} used a multi-channel asymmetric autoencoder to reconstruct IMU signals and classify activities simultaneously. Peng et al.~\cite{peng2018aroma} developed the AROMA model, combining a shared CNN backbone and LSTM to recognize complex activities jointly. Mekruksavanich et al.~\cite{mekruksavanich2025deep} applied a CNN--BiLSTM network for joint prediction of user activity and identity, while Nisar et al.~\cite{nisar2023hierarchical} proposed a hierarchical MTL framework using multi-branch network for activity recognition. Most of these HAR models use hard parameter sharing (shared CNN/LSTM layers with task-specific heads), which allows leveraging relatedness between tasks for better joint modeling of multiple.

Despite progress in modeling, existing university smartphone-sensing datasets are limited in their ability to capture both detailed activity and social context concurrently. Some datasets offer rich activity labels but lack social annotations; others include only coarse social categories or lack detailed, temporally aligned annotations. For example, Bouton-Bessac et al.~\cite{boutonbessac2023daypocketcomplexactivity} collected over 216,000 activity reports from 600+ students across five countries, but did not include social context labels. Aurel et al.~\cite{10.1145/3613904.3642444} surveyed 581 undergraduates using a simple survey but lacked fine-grained activity data. Nathan et al.~\cite{10.1145/3577190.3614129} focused on eating episodes among 678 students, distinguishing only ``eating alone'' versus ``eating with others.'' The DiversityOne dataset~\cite{kammoun2023understanding} spans 782 students in eight countries with 350,000+ half-hour activity and social context diaries, but primarily supports cross-country comparisons and lacks temporally precise, detailed joint annotations.

Our Sedentary and Social Context Dataset (SSCD) addresses these limitations by providing fine-grained, temporally aligned labels for sedentary activities and nuanced social context states. This enables more accurate and integrated modeling of real-world sedentary activity and social context in university environments.

\begin{figure*}[!htbp]
    \centering
    \resizebox{0.9\textwidth}{!}{%
        \includegraphics{images/DySTAN.pdf}%
    }
    \caption{DySTAN architecture for joint classification of sedentary activity and social context. The model processes 13-channel IMU sensor data through shared and task-specific CNN layers, dynamically fuses representations with a cross-stitch unit, applies cross-task attention, and uses bidirectional LSTMs followed by separate output heads for each context.}
    \label{fig:dystan}
\end{figure*}

\section{Methodology}
% Our approach focuses on collecting passive sensor data alongside self-reported labels for sedentary activity and social context. We then leverage this data to develop a multi-task learning model that jointly predicts both contexts.
\subsection{Data Collection}
We developed a custom Android application (\textbf{LogMe}) to collect in-the-wild data using both passive sensors and active self-reports. We process raw IMU sensor data comprising 13 channels: accelerometer (X, Y, Z), gyroscope (X, Y, Z), magnetometer (X, Y, Z), and rotation quaternion (X, Y, Z, W) signals. The app continuously recorded motion and orientation data from these sensors at 50~Hz. Participants received hourly notifications at the 55th minute from 7:00~AM to 9:00~PM. When prompted, users first selected their current \textit{sedentary activity} from the following options: \textbf{Attending Lecture (AL)}, \textbf{Relaxing (R)}, \textbf{Studying (S)}, \textbf{Eating (E)}, or \textbf{Other Activity} (for activities outside these categories). They then reported their \textit{social context}: \textbf{Alone (A)}, \textbf{With Someone (Engaged in Conversation) [WSEIC]}, or \textbf{With Someone (Not Engaged in Conversation) [WSNEIC]}. Each notification remained active for 30 minutes before disappearing if not clicked.

Following institutional ethics approval, participants were recruited via campus email. Sixty students attended a briefing session that outlined the two-week study, explained app usage, and demonstrated notification timing and the app interface. Three students opted out due to privacy concerns, resulting in 57 participants (out of 57 participants, 39 were male and 18 were female, mean age = 20.16 ± 1.92 years ). The study collected 67.2 hours of passive sensor and self-report data focused on fine-grained sedentary activity labels and nuanced social context states, with background data collection minimizing user disruption. Refreshments were provided as an incentive.

\subsection{Pre-Processing}

IMU sensor data were preprocessed to reduce noise and improve signal quality. We explicitly use only raw sensor signals in our analysis \cite{golroudbari2023generalizableendtoenddeeplearning,s22134755}. Accelerometer readings were converted to linear acceleration by removing the gravitational component, then filtered sequentially using a 3rd-order Butterworth high-pass filter (0.3 Hz cutoff)~\cite{napoli2024benchmark} and a low-pass filter (20 Hz cutoff)~\cite{quiroz2017feature} to reduce drift and high-frequency noise. Gyroscope signals were smoothed with a 3rd-order Butterworth low-pass filter at 20 Hz~\cite{quiroz2017feature}, while magnetometer data were denoised using a 2nd-order Chebyshev Type-I low-pass filter with a 0.001 dB passband ripple~\cite{al2022smartphone}. Rotation vector signals~\cite{gayathri2025advancing,he2024human} (quaternions) were used without additional filtering due to their inherent stability reported in prior studies~\cite{moubarak2024smartphone}. After preprocessing, we extracted 1-minute IMU segments aligned with each self-reported response, representing the participant’s activity during that period. The original 50 Hz data were downsampled to 40 Hz to balance computational efficiency and accuracy \cite{cruciani2020feature}. These segments were further divided into overlapping 2.5-second windows with 50\% overlap~\cite{s18061965}, a setup shown effective for activity and context recognition.
\subsection{Proposed Method}

%In this section, we present DySTAN (Dynamic Cross-Stitch with Task Attention Network), a cross-stitch-based architecture with an integrated attention module for multi-task learning. DySTAN leverages shared information between tasks to improve the classification of activities with similar sensor signatures (see Fig \ref{fig:dystan}).

 DySTAN (Dynamic Cross-Stitch with Task Attention Network) shown in Figure \ref{fig:dystan} is a cross-stitch-based architecture with an integrated attention module for multi-task learning. It is designed to jointly classify sedentary activity and social context using smartphone IMU sensors data. Its design allows the model to capture both shared and task-specific temporal features, which is crucial for distinguishing activities with similar sensor signatures while leveraging their underlying commonalities \cite{azadi2024robust, sener2018multi, nisar2023hierarchical, misra2016cross, peng2018aroma}. We process raw IMU sensor data comprising 13 channels for each data window of 100 time steps, capturing rich information on user movement and orientation for classification.

The architecture begins with a shared convolutional feature extractor (1D CNN) consisting of two sequential one-dimensional convolutional layers, each followed by batch normalization and ReLU activation \cite{balaha2023comprehensive, banjarey2022human, ordonez2016deep}. The first convolutional layer (64 filters, kernel size 7) captures broad temporal dependencies across all sensor channels, while the second layer (64 filters, kernel size 5) further refines these features. By sharing this feature extractor, DySTAN efficiently learns temporal patterns common to both tasks, enhancing both data efficiency and model robustness \cite{gu2021survey}. To ensure the model captures distinctions unique to each task, DySTAN branches into two task-specific convolutional streams (Sedentary 1D CNN and Social 1D CNN) following the shared layers. Each stream applies a 1D convolution (128 filters, kernel size 3) with batch normalization and ReLU activation. This design enables each task head to specialize its features, allowing the network to focus on subtle patterns that might be overlooked if all representations were shared \cite{misra2016cross, ruder2017overview,kendall2018multi,sener2018multi}.

\begin{table*}[!t]
\centering
\caption{
Comparison of single-task and multi-task models for joint sedentary activity and social context recognition. Models include CBG , AROMA, DMTLN, METIER, CS (Cross-Stitch), SN (Sluice Network), and DySTAN. Metrics reported are accuracy and macro-F1 for each task and joint accuracy (mean~$\pm$~standard deviation across 5 folds) with best results are bolded.}
\scriptsize
\resizebox{0.9\textwidth}{!}{%
\begin{tabular}{l
    r@{ $\pm$ }l
    r@{ $\pm$ }l
    r@{ $\pm$ }l
    r@{ $\pm$ }l
    r@{ $\pm$ }l
    r@{ $\pm$ }l
    r@{ $\pm$ }l
}
\toprule
\multicolumn{1}{c}{\textbf{Model}} & 
\multicolumn{2}{c}{\textbf{Sedentary Acc.}} & 
\multicolumn{2}{c}{\textbf{Sedentary Macro F1}} & 
\multicolumn{2}{c}{\textbf{Social Acc.}} & 
\multicolumn{2}{c}{\textbf{Social Macro F1}} & 
\multicolumn{2}{c}{\textbf{Joint Acc.}} \\
\cmidrule(lr){2-3} \cmidrule(lr){4-5} \cmidrule(lr){6-7} \cmidrule(lr){8-9} \cmidrule(lr){10-11}
& \multicolumn{2}{c}{$\mu \pm \sigma$}
& \multicolumn{2}{c}{$\mu \pm \sigma$}
& \multicolumn{2}{c}{$\mu \pm \sigma$}
& \multicolumn{2}{c}{$\mu \pm \sigma$}
& \multicolumn{2}{c}{$\mu \pm \sigma$} \\
\midrule

CBG  % CNN-BiLSTM-GRU
& 0.666 & 0.007 & 0.634 & 0.005 & 0.701 & 0.012 & 0.672 & 0.010 & \multicolumn{2}{c}{-} \\
\hdashline[1pt/2pt]

AROMA  
& 0.690 & 0.006 & 0.660 & 0.007 & 0.733 & 0.012 & 0.703 & 0.012 & 0.567 & 0.012 \\

DMTLN 
& 0.650 & 0.011 & 0.621 & 0.011 & 0.696 & 0.011 & 0.663 & 0.008 & 0.512 & 0.015 \\

METIER 
& 0.683 & 0.012 & 0.647 & 0.012 & 0.718 & 0.016 & 0.685 & 0.016 & 0.548 & 0.015 \\

CS  % Cross-Stitch
& 0.668 & 0.032 & 0.641 & 0.033 & 0.679 & 0.026 & 0.655 & 0.029 & 0.512 & 0.013 \\

SN  % Sluice Network
& 0.798 & 0.008 & 0.770 & 0.008 & 0.824 & 0.015 & 0.798 & 0.012 & 0.700 & 0.012 \\

\textbf{DySTAN (ours) }
& \textbf{0.882} & \textbf{0.003} 
& \textbf{0.852} & \textbf{0.004} 
& \textbf{0.902} & \textbf{0.003} 
& \textbf{0.876} & \textbf{0.003} 
& \textbf{0.831} & \textbf{0.005} \\
\bottomrule
\end{tabular}
} % End of resizebox
\label{tab:multitask_results}
\end{table*}

A central innovation of DySTAN is the Dynamic Cross-Stitch Unit (DCSU). Unlike traditional cross-stitch networks that use fixed, static mixing weights~\cite{misra2016cross}, the DCSU dynamically generates mixing weights conditioned on the input. After the task-specific convolutional layers, each feature map is summarized using average pooling and fed into a lightweight controller network. This controller consists of a two-layer multilayer perceptron (MLP): the first layer reduces the pooled feature vector to a compact hidden representation, followed by a second layer that outputs per-channel, per-instance adaptive weights, which are then reshaped to form dynamic mixing matrices. These adaptive weights determine in real time how much information should be exchanged between the two tasks. By enabling input-dependent feature sharing~\cite{sener2018multi,ma2018modeling}, the DCSU provides DySTAN with greater flexibility to adapt to diverse sensor contexts and activity patterns, leading to improved performance in complex, real-world classification settings.

Following dynamic fusion, a Cross-Task Attention module is applied. This module uses multi-head attention to let each task focus on important time steps in the feature sequence of the other task \cite{liu2019end, kharlova2021time}. This mechanism enables DySTAN to capture complex, bidirectional dependencies between sedentary activity and social context, which often influence each other in practice. The use of residual connections ensures that the original task features are preserved, while the attention layer enriches them with additional context \cite{chen2020multi, jou2016deep,khan2021attention}. To further model temporal dependencies, DySTAN applies bidirectional LSTM layers after the attention blocks for each task. These LSTMs (128 units per direction) process the attended features to capture long-range patterns in both forward and backward directions \cite{uddin2024deep,khatun2022deep}. Temporal mean-pooling of the bidirectional LSTM outputs yields fixed-length embeddings that summarize the sequence information for each task.

Finally, DySTAN passes the resulting embeddings through task-specific classification heads. Each head consists of a fully connected layer with ReLU activation, followed by a dropout layer (dropout rate 0.4) for regularization, and then outputs the logits for its respective task. The entire network is trained end-to-end using a combined cross-entropy loss for both tasks and optimized with Adam, ensuring balanced updates to both shared and task-specific parameters for robust multi-task classification. Overall, DySTAN’s architecture—combining shared and task-specific convolutions, dynamic cross-stitch fusion, cross-task attention, and bidirectional LSTMs—is carefully designed to capture complex, overlapping, and distinct temporal patterns required for accurate, joint context classification from raw sensor data.

\subsection{Baselines}
We compare DySTAN against recent state-of-the-art baselines highly relevant to our task. These include single-task CNN-BiLSTM-GRU (CBG) \cite{s21248227}, which integrates convolutional, bidirectional LSTM, and gated recurrent layers, and multi-task models such as AROMA \cite{10.1145/3214277}, DMTLN (Deep Multi-Task Learning Network) \cite{MEKRUKSAVANICH20251350}, and METIER \cite{10.1145/3381012}. Additionally, we evaluate advanced architectures like Cross-Stitch Networks (CN) \cite{misra2016crossstitchnetworksmultitasklearning} and Sluice Networks (SN) \cite{ruder2018latentmultitaskarchitecturelearning} that enable adaptive and dynamic parameter sharing across tasks. All baselines are trained and tested with the same data processing and validation protocols to ensure a fair and rigorous comparison.

\subsection{Implementations Details}

IMU data were segmented into windows of 2.5 seconds with 50\% overlap. All sensor channels were standardized prior to model training, and the labels were appropriately encoded. We employed five-fold stratified cross-validation for evaluation \cite{rabbi2021human}. In each fold, 64\% of the data was used for training, 16\% for validation (taken as 20\% of the training set), and 20\% for testing. Models were trained with the Adam optimizer (learning rate = 1e-3), a batch size of 64, and a maximum of 50 epochs per fold \cite{ordonez2016deep}. Cross-entropy loss was used for both tasks, with class weights calculated from the training data to address class imbalance. Model performance was assessed using accuracy, macro-F1 score, silhouette score (measuring cluster separation and cohesion), intra-class distance (how close samples are within the same class), inter-class distance (how far apart different class centers are), and normalized confusion matrices (visualizing class-wise prediction performance) for both sedentary activity and social context tasks.

\section{Results And Discussion}
%Table~\ref{tab:multitask_results} summarizes the comparative performance of DySTAN and  baselines on the tasks of Sedentary Activity and Social Context classification.

\begin{table*}[!htbp]
\centering
\caption{Ablation study results for DySTAN and its architectural variants on joint classification of Sedentary Activity and Social Context. Ablations include: NSN (No Shared Network), NB (No BiLSTM), and NA (No Attention). Metrics are accuracy, macro-F1, and joint accuracy (mean~$\pm$~standard deviation across 5 folds). Best results per metric are \textbf{bolded}}
\scriptsize
\resizebox{0.9\textwidth}{!}{%
\begin{tabular}{l
    r@{ $\pm$ }l
    r@{ $\pm$ }l
    r@{ $\pm$ }l
    r@{ $\pm$ }l
    r@{ $\pm$ }l
    r@{ $\pm$ }l
    r@{ $\pm$ }l
}
\toprule
\multicolumn{1}{c}{\textbf{Model}} & 
\multicolumn{2}{c}{\textbf{Sedentary Acc.}} & 
\multicolumn{2}{c}{\textbf{Sedentary Macro F1}} & 
\multicolumn{2}{c}{\textbf{Social Acc.}} & 
\multicolumn{2}{c}{\textbf{Social Macro F1}} & 
\multicolumn{2}{c}{\textbf{Joint Acc.}} \\
\cmidrule(lr){2-3} \cmidrule(lr){4-5} \cmidrule(lr){6-7} \cmidrule(lr){8-9} \cmidrule(lr){10-11}
& \multicolumn{2}{c}{$\mu \pm \sigma$}
& \multicolumn{2}{c}{$\mu \pm \sigma$}
& \multicolumn{2}{c}{$\mu \pm \sigma$}
& \multicolumn{2}{c}{$\mu \pm \sigma$}
& \multicolumn{2}{c}{$\mu \pm \sigma$} \\
\midrule

DySTAN
& \textbf{0.882} & \textbf{0.003} 
& \textbf{0.852} & \textbf{0.004} 
& \textbf{0.902} & \textbf{0.003} 
& \textbf{0.876} & \textbf{0.003} 
& \textbf{0.831} & \textbf{0.005} \\

NSN 
& 0.839 & 0.004 
& 0.797 & 0.005 
& 0.866 & 0.001 
& 0.826 & 0.002 
& 0.766 & 0.004 \\

NB
& 0.788 & 0.003 
& 0.739 & 0.005 
& 0.819 & 0.001 
& 0.751 & 0.003 
& 0.687 & 0.003 \\

NA
& 0.874 & 0.001 
& 0.840 & 0.003 
& 0.896 & 0.001 
& 0.867 & 0.001 
& 0.817 & 0.003 \\

\bottomrule
\end{tabular}
} % End of resizebox

\label{tab:ablation_results_double}
\end{table*}

DySTAN outperforms (see Table~\ref{tab:multitask_results}) all baselines across every major metric, achieving a mean accuracy of 0.882$\pm$0.003 for Sedentary Activity and 0.902$\pm$0.003 for Social Context. This represents a relative improvement of 10.5\% in Sedentary Activity accuracy and 9.5\% in Social Context accuracy compared to the strongest multi-task baseline, Sluice Network (0.798 and 0.824, respectively). Macro F1 scores exhibit similar trends: DySTAN achieves 0.852$\pm$0.004 for Sedentary Activity and 0.876$\pm$0.003 for Social Context, surpassing all baseline models. The joint accuracy metric, reflecting the model’s ability to correctly predict both Sedentary Activity and Social Context simultaneously, further highlights DySTAN’s advantage: DySTAN reaches a joint accuracy of 0.831$\pm$0.005, a 19\% increase over Sluice Network (0.700$\pm$0.012). Other multi-task frameworks, such as DMTLN and METIER, yield much lower joint accuracies (0.512 and 0.548), underscoring the additional benefit provided by DySTAN’s dynamic feature-sharing and attention mechanisms.

The single-task CBG baseline performs notably worse than all multi-task frameworks, with only 0.666$\pm$0.007 accuracy on Sedentary Activity and 0.701$\pm$0.012 on Social Context. DySTAN outperforms CBG by 32.5\% in Sedentary Activity accuracy and 28.7\% in Social Context accuracy, illustrating the clear advantage of multi-task learning and shared representation. Cross-Stitch and Sluice Network models, which use cross-stitch-based parameter sharing, generally surpass conventional multi-task models (AROMA, DMTLN, METIER), but still fall short of DySTAN. Sluice Network, the strongest of these, achieves 0.798$\pm$0.008 (Sedentary Activity), 0.824$\pm$0.015 (Social Context), and 0.700$\pm$0.012 (joint accuracy).

Macro F1 results closely mirror the accuracy metrics, confirming that DySTAN not only achieves high overall accuracy but also maintains balanced performance across all classes, including underrepresented categories. For example, DySTAN’s macro F1 for Sedentary Activity is 10.6\% higher than Sluice Network (0.852$\pm$0.004 vs. 0.770$\pm$0.008), and 29.1\% higher than the best conventional multi-task baseline, AROMA (0.660$\pm$0.007).

Another important observation is the low standard deviation achieved by DySTAN across all metrics (typically $\pm$0.003–0.005), indicating highly consistent performance across cross-validation folds. This stability contrasts with higher variance observed in baselines such as Cross-Stitch (Sedentary Activity accuracy SD = 0.032), highlighting that DySTAN not only achieves superior accuracy but is also more robust and reliable across different participant splits and real-world data variations.

\subsection{Ablation Study: Component-Wise Impact in DySTAN}
The ablation results in Table~\ref{tab:ablation_results_double} highlight the incremental benefit of each architectural component in DySTAN for joint Sedentary Activity and Social Context classification. Removing the shared convolutional layers (``No Shared Network'') results in a drop in all metrics (e.g., Joint Accuracy falls to 0.766±0.004), emphasizing the importance of learning shared temporal patterns for effective multi-tasking. Omitting bidirectional LSTM layers (``No BiLSTM'') causes the largest decline (Joint Accuracy: 0.687±0.003), showing that modeling long-range temporal dependencies is essential for context recognition. Excluding the cross-task attention mechanism (``No Attention'') also degrades joint and macro-F1 scores, confirming that attention-based information sharing improves the model’s ability to leverage task interdependencies.

Across all ablations, the full DySTAN architecture consistently delivers the highest and most stable performance (Sedentary Activity Accuracy: 0.882±0.003, Social Context Accuracy: 0.902±0.003, Joint Accuracy: 0.831±0.005), underscoring that shared convolutions, dynamic cross-stitch fusion, attention, and BiLSTM layers are all critical and complementary for state-of-the-art results in multi-context recognition from sensor data.
\begin{figure*}[htbp]
    \centering
    \includegraphics[width=0.9\textwidth]{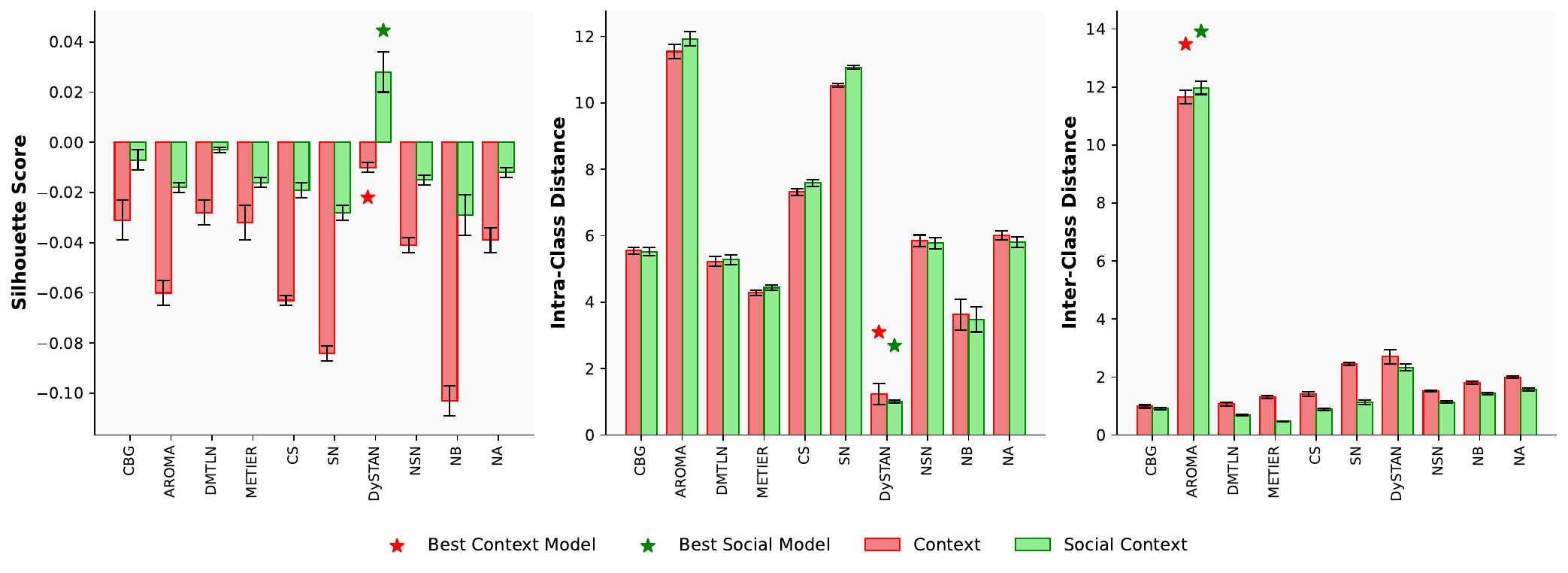}
    \caption{Comparison of Context and Social Context metrics across models. Stars indicate best performing models.}
    \label{fig:metric_comparison}
\end{figure*}

\begin{figure*}[!htbp]  % 'figure*' spans the figure across both columns
    \centering
    \includegraphics[width=0.9\textwidth]{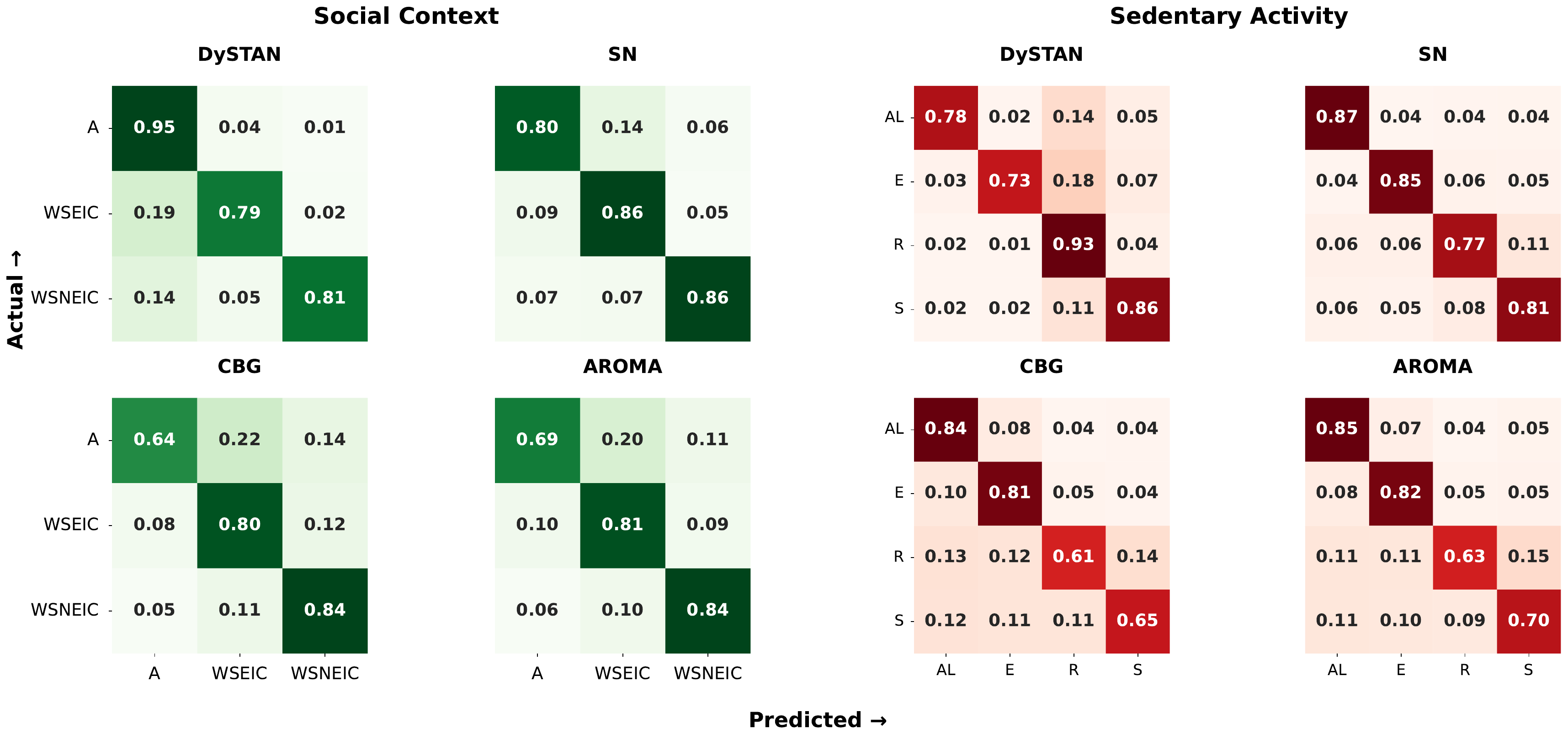}  % Adjust the path as needed
    \caption{Confusion matrices for Social Context and Sedentary Activity classification are presented for each best model (DySTAN, SN, CBG, AROMA). For Social Context, the classes are Alone (A), With Someone (Engage in Conversation) (WSEIC), and With Someone (Not Engage in Conversation) (WSNEIC). The Sedentary Activity classes include Attending Lectures (AL), Eating (E), Relaxing (R), and Studying (S).}
    \label{fig:confusion_matrices}
\end{figure*}

\subsection{Metric-Based Analysis: Cluster Quality and Class Separability}

Figure~\ref{fig:metric_comparison} visualizes three key metrics—Silhouette Score, Intra-Class Distance, and Inter-Class Distance—for all models and ablation variants.
DySTAN produces well-separated and compact class clusters in the learned embedding space, as shown by a high silhouette score, low intra-class distance, and high inter-class distance in the figure, resulting in superior feature representation and class discriminability. DySTAN achieves the highest silhouette score and inter-class distance while maintaining one of the lowest intra-class distances for both Sedentary Activity and Social Context tasks. Notably, classic multi-task models (AROMA, DMTLN, METIER) and single-task CBG exhibit lower silhouette scores and higher intra-class distances, suggesting overlapping or dispersed clusters and inferior separation between classes. Models with cross-stitch sharing (CS, SN) improve upon these metrics, but still lag behind DySTAN—especially in inter-class distance, which is critical for robust classification. Among ablation variants, removing shared representation (NSN), attention (NA), or LSTM layers (NB) all degrade the cluster quality, with reductions in silhouette and inter-class distance, as well as increased intra-class spread. These results underscore the importance of each DySTAN component in achieving tight, distinct, and well-structured class clusters.

In summary, DySTAN’s architecture delivers the best trade-off between compactness (intra-class) and separability (inter-class), directly translating to higher classification accuracy and robustness in real-world multi-task context recognition.
\subsection{Model Wise Confusion Matrix Insights}

The confusion matrices shown in Fig.~\ref{fig:confusion_matrices} demonstrate that DySTAN achieves the highest true positive rates for all classes in both Social Context and Sedentary Activity, with minimal confusion between similar classes. For example, DySTAN correctly classifies the `Alone' (A) class 95\% of the time and `With Someone (Engaged In Conversation)' [WSEIC] 79\% of the time, outperforming Sluice Network (SN), CNN-BiLSTM-GRU (CBG), and AROMA, which all display more off-diagonal errors. For Sedentary Activity, DySTAN also demonstrates strong diagonal dominance, with `Relaxing' (R) and `Studying' (S) recognized at 93\% and 86\% respectively, while `Attending Lecture' (AL) and `Eating' (E) are recognized at 78\% and 73\%. In contrast, baseline models like CBG and AROMA show more substantial misclassification between activities such as `Relaxing' and `Studying'. Overall, DySTAN's confusion matrices indicate more precise, consistent predictions with fewer confusions between closely related contexts, validating its effectiveness for fine-grained context recognition.

\section{Conclusion and Limitations}
This paper presented DySTAN, a neural model that combines dynamic cross-stitch units and cross-task attention for joint recognition of sedentary activities and social contexts from smartphone IMU data. Unlike standard Cross-Stitch Networks, DySTAN dynamically adapts feature sharing per instance and leverages cross-task temporal attention, leading to consistently higher accuracy and macro-F1 scores across all classes on student data. Ablation results confirm that these components drive DySTAN’s improvements over strong multi-task baselines, demonstrating the value of adaptive feature fusion and task interaction for context recognition in naturalistic settings.

A key limitation of this study is the demographic scope of the dataset, which was drawn exclusively from a university student population.  Future work will explore expanding data collection to more diverse populations and settings.
\bibliographystyle{ACM-Reference-Format}
\bibliography{sample-base}
\end{document}